\documentclass[prl,twocolumn,preprintnumbers]{revtex4}
\usepackage{amsmath,amssymb,bm,graphicx}

\newcommand\grad{\bm{\nabla}}

\renewcommand\>{{\rangle}}
\newcommand\x{{\bm{x}}}
\newcommand\y{{\bm{y}}}
\newcommand\p{{\bm{p}}}
\newcommand\q{{\bm{q}}}
\newcommand\kFB{k_{\mathrm{F}\!B}}
\newcommand\eff{{\mathrm{eff}}}
\newcommand\cri{{\mathrm{cri}}}
\newcommand{\sect}[1]{{\em#1.}---}
\newcommand{\dif}{\mathrm{d}}
\newcommand{\e}{\mathrm{e}}
\newcommand{\I}{\mathrm{i}}

\begin{document}
\preprint{INT-PUB 08-11}

\title{Universal Fermi gases in mixed dimensions}
\author{Yusuke~Nishida}
\author{Shina~Tan}
\affiliation{Institute for Nuclear Theory, University of Washington,
             Seattle, Washington 98195-1550, USA}

\begin{abstract}
 We investigate a two-species Fermi gas in which one species is confined
 in a two-dimensional plane (2D) or one-dimensional line (1D) while the
 other is free in the three-dimensional space (3D).  We discuss the
 realization of such a system with the interspecies interaction tuned to
 resonance.  When the mass ratio is in the range
 $0.0351<m_\mathrm{2D}/m_\mathrm{3D}<6.35$ for the 2D-3D mixture or
 $0.00646<m_\mathrm{1D}/m_\mathrm{3D}<2.06$ for the 1D-3D mixture, the
 resulting system is stable against the Efimov effect and has universal
 properties.  We calculate key quantities in the many-body phase
 diagram.  Other possible scale-invariant systems with short-range
 few-body interactions are also elucidated.
\end{abstract}

\date{June 2008}


\maketitle

\sect{Introduction}
Experiments using ultracold atomic gases have attracted considerable
interest because of their high designability and tunability.  Not only
can the strength of interaction be varied via the Feshbach resonance,
but also the dimensionality of space can be changed by means of strong
optical lattices.  The Bose-Einstein condensation in trapped atoms has
been realized both in one and two dimensions~\cite{1d-2d_BEC}.  A strongly
interacting one-dimensional Fermi gas has been observed~\cite{1d_Fermi},
which is an atomic realization of Tomonaga-Luttinger liquid.

What the cold atom community has not paid attention to is the system in
mixed dimensions, namely, the system where different types of particles
live in different spatial dimensions.  Actually the idea of mixed
dimensions is commonly used in various subfields of physics.  For
example, in the brane world model of the Universe, ordinary matter
is confined in a three-dimensional space embedded in higher dimensions
where gravity can propagate~\cite{Maartens:2003tw}.  Also, recently
realized graphene can be regarded as a system in mixed dimensions where
photons induce the three-dimensional Coulomb interaction between
electrons confined in a two-dimensional sheet~\cite{graphene}.

In this Letter, we discuss the realization of an analogous system using
cold atoms with two fermionic species $A$ and $B$, where $A$ atoms are
confined in a two-dimensional plane (2D) or one-dimensional line (1D)
while $B$ atoms are free in the three-dimensional space (3D).  At low
energies, we show that the interspecies interaction in a certain range
of mass ratio is characterized by a single parameter, the effective
scattering length $a_\eff$, whose value is arbitrarily tunable by
current experiments.  In particular, if the effective interaction is
tuned to resonance $|a_\eff|=\infty$, such a system is scale invariant
and has universal properties.  (Here intra-species interactions are
neglected and we use the term ``universal'' in the narrow sense that the
interaction does not have any dimensionful parameters.)  The
corresponding many-body system forms a novel type of unitary Fermi gas
in mixed dimensions.

\sect{Experimental realization}
To confine only $A$ atoms, we note that the strength of the optical trap
experienced by neutral atoms is proportional to the polarizability which
depends on the atomic species and the laser frequency $\omega_L$. By
tuning $\omega_L$ at a zero of the polarizability of $B$ atoms, which
exists between any two nearest resonances, we can confine only $A$
atoms.  If the optical trap with such a laser frequency is
$d_\perp$ dimensional with $d_\perp=1$ or $2$, $A$ atoms are confined in
2D or 1D, respectively, while $B$ atoms are free from the confinement.

If the confinement potential applied to $A$ atoms is harmonic with the
oscillator frequency $\omega_\perp$, an $A$ atom interacting with a $B$
atom is described by the Hamiltonian:
$H = -\frac{\hbar^2\grad_{\x_A}^{\,2}}{2m_A}
 + \frac12m_A\omega_\perp^{\,2}\x_{A\perp}^{\,2}
 - \frac{\hbar^2\grad_{\x_B}^{\,2}}{2m_B} + V(\x_A,\x_B)$.
We introduce a notation $\x_A=(\x_{A\parallel},\x_{A\perp})$, where
$\x_{A\perp}$ is the $d_\perp$-dimensional coordinates affected by the
confinement potential. $V(\x_A,\x_B)$ is the bare interspecies interaction.
In the limit of zero-range interaction, it is characterized by a single
parameter, the $s$-wave scattering length $a$, which is arbitrarily
tunable by means of the Feshbach resonance.

Because of the partial confinement of the $A$ atom, the scattering
property between $A$ and $B$ atoms is modified from the free-space
scattering.  In particular, an infinite number of confinement-induced
resonances appears.  To determine the position of resonances as a
function of $l_\perp/a$ with
$l_\perp\equiv\sqrt{\frac{\hbar}{m_A\omega_\perp}}$, we solve the
Schr\"odinger equation $H\Psi(\x_A,\x_B)=E\Psi(\x_A,\x_B)$ in the
zero-energy limit $E-\frac{d_\perp}2\hbar\omega_\perp\to0$.  After
excluding the center-of-mass motion in the parallel direction, the
solution is written as
\begin{align}
 \Psi &= \int\dif^{d_\perp}\!\y_\perp
 G(\x_{A\parallel}-\x_{B\parallel},\x_{A\perp},\x_{B\perp};
 \bm0,\y_\perp,\y_\perp) f(y_\perp) \notag\\
 &\quad + C\phi_0(x_{A\perp}),
\end{align}
where
$\phi_0(x_{A\perp})=\exp\bigl[-x_{A\perp}^{\,2}/(2l_\perp^{\,2})\bigr]$
is the ground state wave function of the $d_\perp$-dimensional harmonic
oscillator.  $G$ is the Green's function given by 
\begin{equation}
 \begin{split}
  G &= \int_0^\infty\!\frac{\dif\tau}{\tau^{3/2}}
  \left(\frac{\e^\tau}{\sinh\tau}\right)^{d_\perp/2}\exp\!
  \left[-\frac{(\x_{A\parallel}{-}\x_{B\parallel})^2}{2l_\perp^{\,2}(u+1)\tau}
  \right. \\ &\left.
  -\frac{(x_{A\perp}^{\,2}{+}y_\perp^{\,2})\cosh\tau{-}2\x_{A\perp}\cdot\y_\perp}
  {2l_\perp^{\,2}\sinh\tau}
  -\frac{(\x_{B\perp}{-}\y_\perp)^2}{2l_\perp^{\,2}u\tau}\right]
 \end{split}
\end{equation}
with $u\equiv m_A/m_B$ being the mass ratio.  The unknown function
$f(y_\perp)/C$ is uniquely determined by imposing the short-range
boundary condition
$\Psi(\x_A,\x_B)\propto\frac1{|\x_A-\x_B|}-\frac1a+O(|\x_A-\x_B|)$ when
$|\x_A-\x_B|\to0$~\cite{Massignan:2006}.  The asymptotic form at a large
separation $|\x_A-\x_B|\to\infty$,
\begin{equation}\label{eq:Psi_AB}
 \Psi(\x_A,\x_B) \propto \left[\frac1{\mathcal{D}(\x_A,\x_B)}
			  -\frac1{a_\eff}\right]\,\phi_0(x_{A\perp}),
\end{equation}
gives an effective scattering length $a_\eff$, and whenever $a_\eff$
diverges, a resonance occurs.  Here 
$\mathcal{D}(\x_A,\x_B)\equiv\sqrt{(\x_{A\parallel}-\x_{B\parallel})^2
+\frac{m_B}{m_{AB}}\x_{B\perp}^{\,2}}$
with $m_{AB}=\frac{m_Am_B}{m_A+m_B}$ being the reduced mass is the
``interparticle distance'' with the anisotropic weights to the parallel
and perpendicular directions.  

\begin{figure}[tp]
 \includegraphics[width=0.24\textwidth,clip]{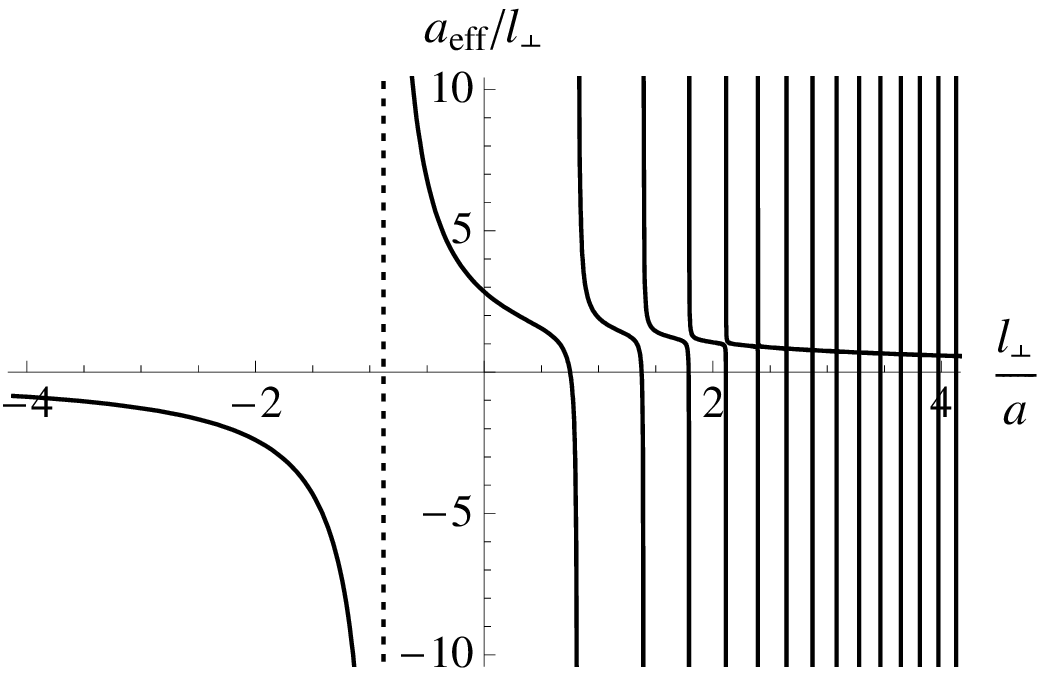}\hfill
 \includegraphics[width=0.24\textwidth,clip]{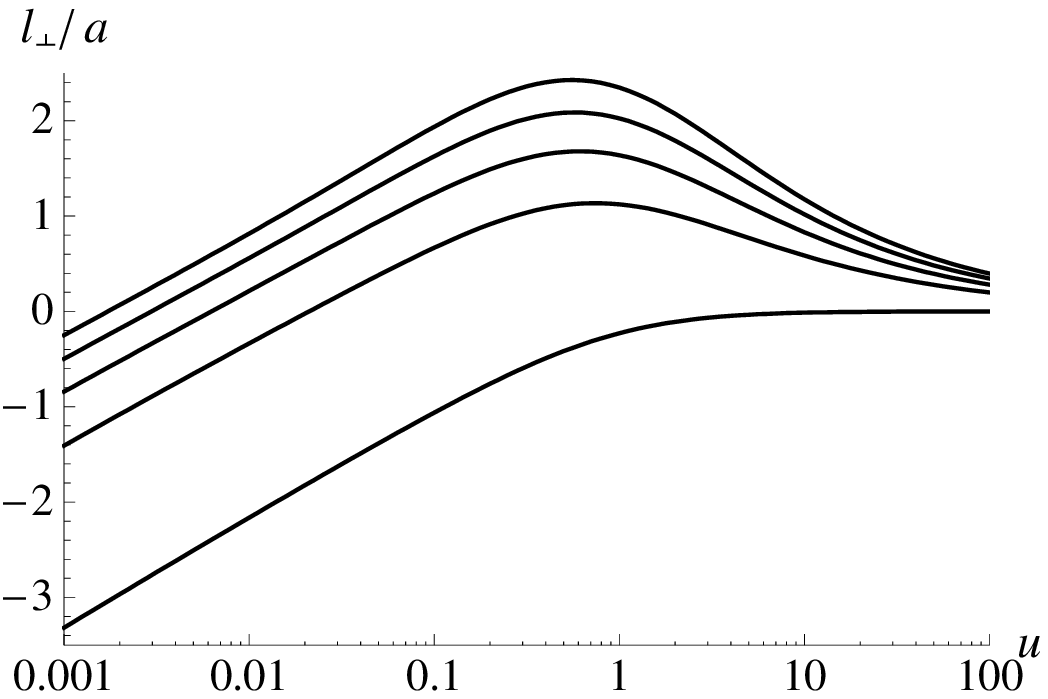}
 \caption{(left) $a_\eff/l_\perp$ as a function of $l_\perp/a$ for
 $d_\perp=1$ (2D-3D mixture).  The mass ratio is chosen to be $u=0.15$.
 The vertical dotted line indicates the position of the broadest
 resonance.  (right) Lowest five values of $l_\perp/a$ to achieve the
 resonances $|a_\eff|/l_\perp=\infty$ as functions of $u$.
 \label{fig:2D3D}}
\end{figure}

\begin{figure}[tp]
 \includegraphics[width=0.24\textwidth,clip]{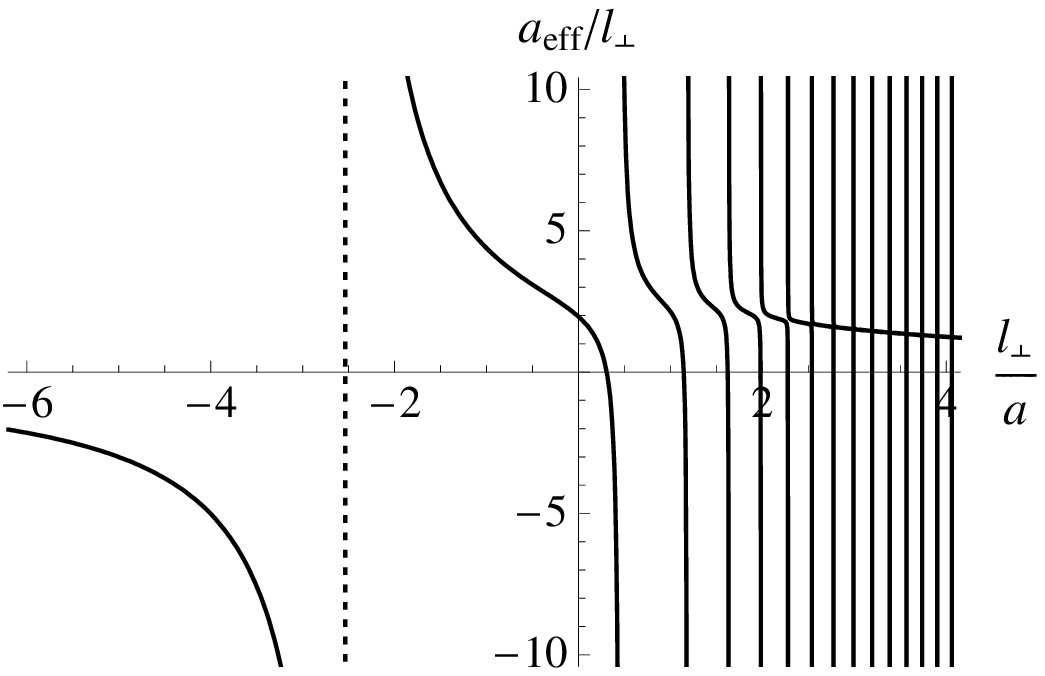}\hfill
 \includegraphics[width=0.24\textwidth,clip]{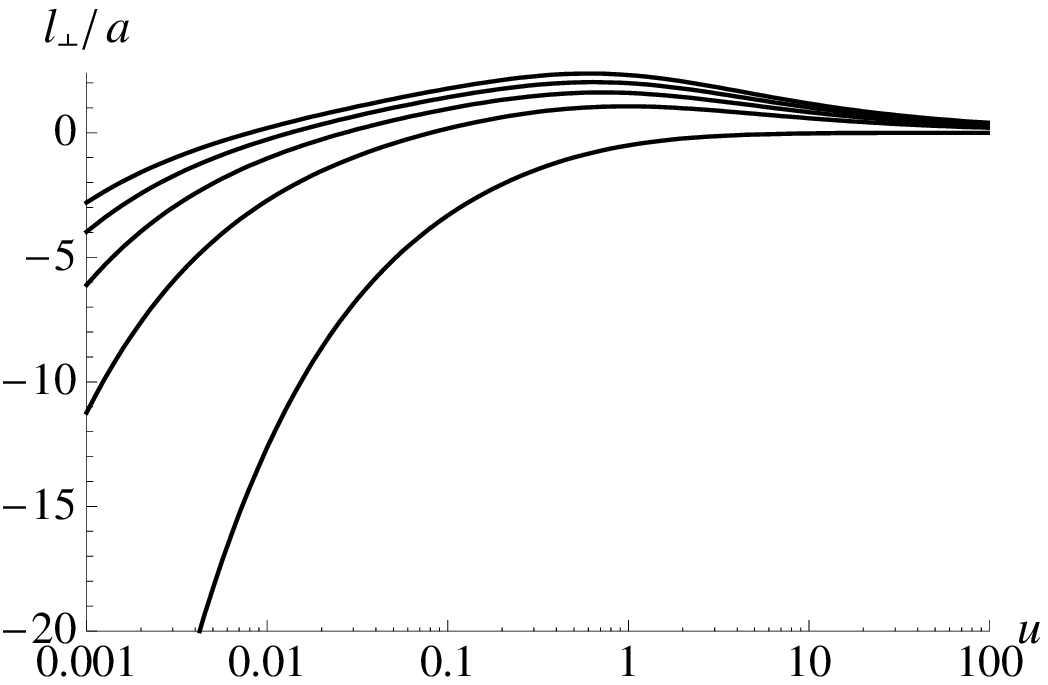}
 \caption{Same as Fig.~\ref{fig:2D3D} for $d_\perp=2$ (1D-3D mixture).
 \label{fig:1D3D}}
\end{figure}

Numerically obtained $a_\eff/l_\perp$ as a function of $l_\perp/a$ is
plotted for $d_\perp=1$ (2D-3D mixture) in Fig.~\ref{fig:2D3D} and for
$d_\perp=2$ (1D-3D mixture) in Fig.~\ref{fig:1D3D}.  We chose the mass
ratio $u=0.15$ corresponding to the physical case of
$A={}^6\mathrm{Li}$ and $B={}^{40}\mathrm{K}$.  These figures show
that $a_\eff$ can be tuned to any desired value by varying $a$
or $l_\perp$.  In particular, the values of $l_\perp/a$ to achieve the
resonances $|a_\eff|/l_\perp=\infty$ are plotted as functions of $u$ in
Figs.~\ref{fig:2D3D} and \ref{fig:1D3D}.

If the confinement length $l_\perp$ is much smaller than any other
length scales of the system such as $a_\eff$ and a mean interparticle
distance at finite density, one can neglect the motion of $A$ atoms in
the confinement direction.  Consequently, the resulting system is
mixed-dimensional, namely, $A$ atoms are confined in 2D or 1D while $B$
atoms are in 3D. The interspecies interaction is solely characterized by
the effective scattering length $a_\eff$ defined in
Eq.~\eqref{eq:Psi_AB}.  When $a_\eff>0$, there is a shallow bound state
with the binding energy
$E_\text{binding}=-\hbar^2/(2m_{AB}a_\eff^{\,2})$.

If the effective interaction is tuned to resonance $|a_\eff|=\infty$,
the system is scale invariant and universal, i.e., independent of the
short-range physics.  The corresponding many-body system forms a novel
type of unitary Fermi gas in mixed dimensions.  In the rest of this
Letter, we concentrate on the system at the resonance. 

\sect{Stability of the system at resonance}
An important question regarding our novel system is its stability.  In
the usual 3D case, it is known that two heavy and one light fermions
with mass ratio greater than $13.6$ develop deep bound states in the
limit of zero-range interaction (Efimov effect)~\cite{Efimov:1972}.
Accordingly the corresponding many-body system is not stable toward
collapse.  To establish the stability of the unitary Fermi gas
in mixed dimensions, we study three-body problems and determine the
range of $u$ where the Efimov effect is absent.

We first consider the problem of two $A$ atoms interacting with one $B$
atom.  The wave function $\Psi(\x_{A1},\x_{A2},\x_B)$ for
$\mathcal{D}(\x_{Ai},\x_B)>0$ satisfies the Schr\"odinger equation:
$\Bigl[-\frac{\hbar^2\grad_{\x_{A1}}^{\,2}}{2m_A}
 - \frac{\hbar^2\grad_{\x_{A2}}^{\,2}}{2m_A}
 - \frac{\hbar^2\grad_{\x_B}^{\,2}}{2m_B}\Bigr]\Psi = E\Psi$.
Now $\x_A=\x_{A\parallel}$ is two- or one-dimensional coordinates.  The
resonant interspecies interaction is taken into account by the
short-range boundary condition:
$\Psi\propto\mathcal{D}(\x_{Ai},\x_B)^{-1}+O[\mathcal{D}(\x_{Ai},\x_B)]$
when $\mathcal{D}(\x_{Ai},\x_B)\to0$ [see Eq.~\eqref{eq:Psi_AB}].  In
hyperspherical coordinates, the wave function at a short distance
behaves as~\cite{exact_relationship}
\begin{equation}\label{eq:Psi_AAB}
 \Psi(\x_{A1},\x_{A2},\x_B)\simeq R^\gamma \psi_l(\Omega),
\end{equation}
where $R$ is the hyperradius and $\Omega$ denotes the hyperangular
variables.  $\gamma$ is the scaling exponent of the wave function
classified by the quantum number $l$ and can be determined by writing
$\Psi$ in terms of the Green's function and imposing the short-range
boundary condition~\cite{Petrov:2003}.


In the 2D-3D mixture, $l=0,\pm1,\pm2,\ldots$ is the orbital angular
momentum projected to the 2D plane and $\gamma$ satisfies the
equation~\cite{footnote_2D}
\begin{equation}\label{eq:gamma_2D}
 \begin{split}
  & -\frac{\sqrt{2u+1}}{u+1} 
  = \int_0^\pi\frac{\dif\theta}{\pi}\cos(l\theta)
  \Bigl[g\Bigl(\gamma,\arccos\frac{u\cos\theta}{u+1}\Bigr) \\
  &\mspace{165mu}
  + g\Bigl(-\gamma-3,\arccos\frac{u\cos\theta}{u+1}\Bigr)\Bigr],
 \end{split}
\end{equation}
with
\begin{equation}
 \begin{split}
  & g(\gamma,\beta) \equiv \pi^{-1/2}\,\Gamma(-\gamma-1)\,
  \Gamma\bigl(\gamma+\frac{3}{2}\bigr)\,\e^{-\I\beta(\gamma+1)} \\
  &\mspace{110mu} \times{}_2\mspace{-1mu}F_1
  \Bigl(\frac{1}{2},-\gamma-1;-\gamma-\frac{1}{2};\e^{2\I\beta}\Bigr).
 \end{split}
\end{equation}
Here ${}_2\mspace{-1mu}F_1$ is the hypergeometric function.
Equation~\eqref{eq:gamma_2D} has a pair of solutions, $\gamma_+$ and
$\gamma_-$, related to each other by $\gamma_++\gamma_-=-3$.   For the
$p$-wave channel $|l|=1$, as the mass ratio $u$ increases from $0$ to
the critical value $u_{\max}=6.35$, 
$\gamma_+$ decreases from $0$ to $-\frac32$.  When $u$ is increased
further, $u>u_{\max}$, $\gamma_\pm$ become complex conjugates
$\gamma_\pm=-\frac32\pm is_0$, indicating that the wave function
\eqref{eq:Psi_AAB} oscillates as $R^{-3/2}\sin(s_0\ln R+\delta)$ and the
Efimov effect takes place.  The critical mass ratio for other odd $l$ is
greater than $6.35$ and the Efimov effect is absent for any even $l$.

On the other hand, in the 1D-3D mixture, $l=\pm1$ in
Eq.~\eqref{eq:Psi_AAB} corresponds to the parity of the wave function
and $\gamma$ satisfies the equation~\cite{footnote_1D}
\begin{equation}\label{eq:gamma_1D}
  \frac{\sqrt{2u+1}}{u+1}
  = \frac{\cos\bigl[\left(\gamma+1\right)\alpha\bigr]
  + l\cos\bigl[\left(\gamma+1\right)\left(\pi-\alpha\right)\bigr]}
  {\left(\gamma+1\right)\sin\left[\pi\left(\gamma+1\right)\right]}
\end{equation}
with $\alpha\equiv\arccos\frac{u}{u+1}$.  Similarly, this equation has a
pair of solutions, $\gamma_+$ and $\gamma_-$, related to each other by
$\gamma_++\gamma_-=-2$.  For the odd-parity channel $l=-1$, when $u$
exceeds the critical value $u_{\max}=2.06$, 
$\gamma_\pm$ become complex conjugates $\gamma_\pm=-1\pm is_0$,
indicating the Efimov effect. In the even-parity channel, the Efimov
effect does not take place.

We then consider the problem of one $A$ atom interacting with two $B$
atoms.  The wave function 
$\Psi(\x_A,\x_{B1},\x_{B2})\simeq R^\gamma\psi_l(\Omega)$
satisfies the Schr\"odinger equation and the short-range boundary
condition analogous to the previous case.  
$\gamma$ in this case satisfies an integral
equation~\cite{exact_relationship}.   For the $p$-wave or odd-parity
channel, we obtain
\begin{equation}
 -S_\p = \!\int\!\frac{\dif^3\q}{2\pi^2}
  \frac{\hat\p_\parallel\cdot\hat\q_\parallel}
  {p^2+\frac2{u+1}\p_\parallel\cdot\q_\parallel+q^2}
  \frac{(q/p)^{\gamma+1}}{\sqrt{q^2{-}\frac1{(u+1)^2}q_\parallel^{\,2}}}
  S_\q,
 \end{equation}
where $\p=(\p_\parallel,\p_\perp)$ and 
$S_\p\equiv S\left(|\p_\perp|/|\p_\parallel|\right)$ is an unknown
function.  There is a pair of solutions, $\gamma_+$ and $\gamma_-$,
related to each other by $\gamma_++\gamma_-=-4$.  As the mass ratio $u$
decreases from $\infty$ to the critical value $u_{\min}$, $\gamma_+$
decreases from $0$ to $-2$.  When $u$ is decreased further, $u<u_{\min}$,
$\gamma_\pm$ become complex conjugates $\gamma_\pm=-2\pm is_0$,
indicating the Efimov effect.  The critical mass ratios are found to be
$u_{\min}=0.0351$ 
for the 2D-3D mixture and $u_{\min}=0.00646$ 
for the 1D-3D mixture.  In other channels, the Efimov effect does not
take place for the mass ratio greater than $u_{\min}$. 

So far we have supposed that both atomic species $A$ and $B$ are
fermionic.  If either $A$ or $B$ atoms are bosonic, one can confirm that
the Efimov effect takes place for any mass ratio in the $s$-wave or
even-parity channel.  Therefore, for the stability of many-body systems,
both atomic species have to be fermionic with the mass ratio between
$u_{\min}$ and $u_{\max}$.  For example, the combination of fermionic
atoms, $A={}^6\mathrm{Li}$ and $B={}^{40}\mathrm{K}$ ($u=0.15$), can be
used to realize both 2D-3D and 1D-3D mixtures, while the opposite
combination, $A={}^{40}\mathrm{K}$ and $B={}^6\mathrm{Li}$ ($u=6.67$),
suffers the Efimov effect in both mixtures.

\begin{figure}[tp]
 \includegraphics[scale=0.9,clip]{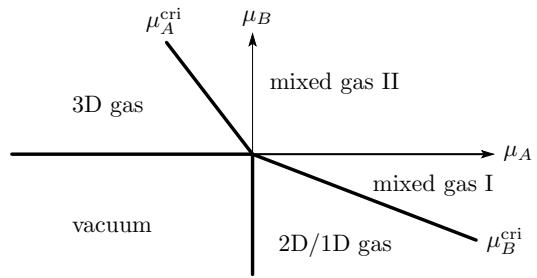}
 \caption{Phase diagram in the plane of $\mu_A$ and $\mu_B$ at zero
 temperature.  \label{fig:phase_diagram}} 
\end{figure}

\sect{Many-body physics}
We now proceed to the many-body physics of unitary Fermi gas in mixed
dimensions.  It is convenient to work in the grand-canonical ensemble by
introducing chemical potentials $\mu_A$ to $A$ atoms and $\mu_B$ to $B$
atoms.  The phase diagram of the system in the plane of $\mu_A$ and
$\mu_B$ is depicted in Fig.~\ref{fig:phase_diagram}.  There are four
distinct regions: vacuum ($\mu_A<0,\,\mu_B<0$), pure 2D or 1D gas
($\mu_A>0,\,\mu_B<\mu_B^\cri$), pure 3D gas
($\mu_B>0,\,\mu_A<\mu_A^\cri$), and mixed gas (rest of the phase
diagram).  As we discuss below, the mixed-gas phase can be divided into
two phases: mixed gas I ($\mu_B<0$) and mixed gas II ($\mu_B>0$).  It is
possible that these mixed-gas phases are further divided into several
phases by nontrivial many-body physics including inter- and
intra-species pairings.

Because of the lack of translational symmetry in the perpendicular
direction, the density of $B$ atoms $n_B(|\x_\perp|)$ is a nontrivial
function of the distance from the 2D plane or 1D line.  If $\mu_B<0$
(mixed gas I), all $B$ atoms are localized around the 2D plane or 1D
line because they are attracted by $A$ atoms.  Accordingly $B$ atoms
have a vanishing density at infinity $n_B(|\x_\perp|\to\infty)\to0$.
When $\mu_B$ exceeds zero (mixed gas II), a portion of $B$ atoms are no
longer bound due to the Fermi pressure.  In this phase, the density of
$B$ atoms at infinity is given by that of noninteracting fermions
$n_B(|\x_\perp|\to\infty)\to\frac{\left(2m_B\mu_B\right)^{3/2}}{6\pi^2\hbar^3}$.

\begin{figure}[tp]
 \includegraphics[width=0.24\textwidth,clip]{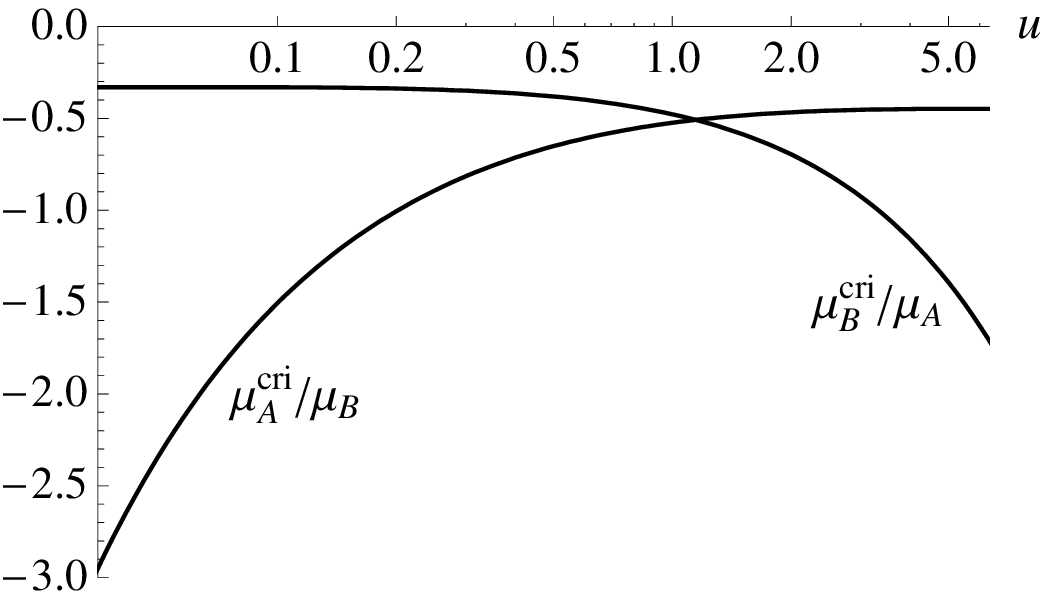}\hfill
 \includegraphics[width=0.24\textwidth,clip]{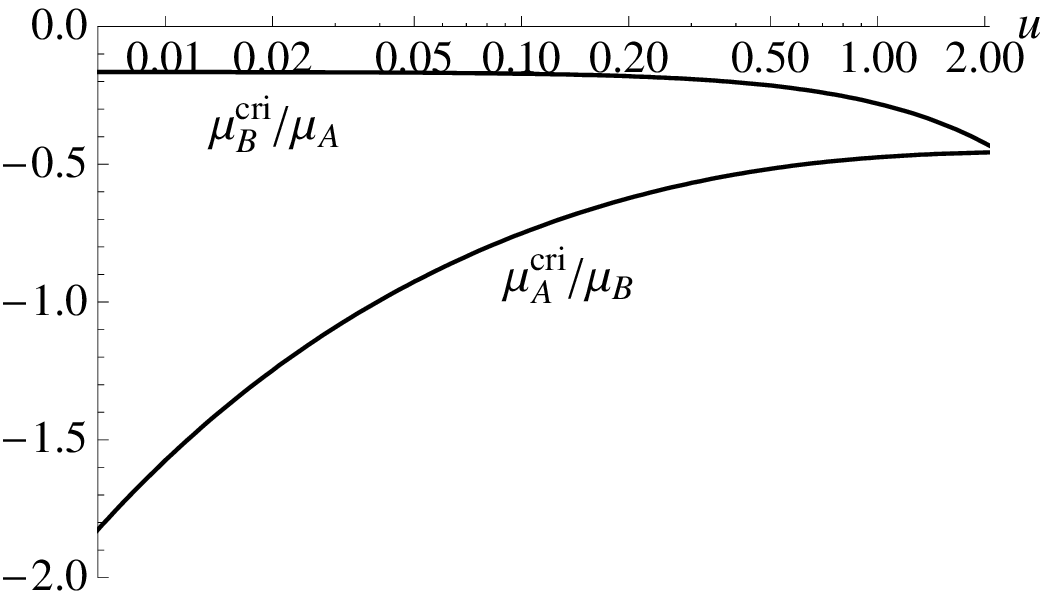}
 \caption{$\mu_A^\cri/\mu_B$ and $\mu_B^\cri/\mu_A$ as functions of $u$
 for the 2D-3D mixture (left) and 1D-3D mixture (right).
 \label{fig:chemical_potential}}
\end{figure}

In order to make the discussion more quantitative, we estimate the key
parameters in the phase diagram.  The chemical potential at the phase
boundary $\mu_{A(B)}^\cri$ normalized by the Fermi energy of majority
atoms $\mu_{B(A)}$ (corresponding to the slope in
Fig.~\ref{fig:phase_diagram}) is a universal number because of the scale
invariance of the interaction.  To calculate
$\mu_A^\cri/\mu_B$, we employ a many-body variational wave function
assuming single particle-hole excitations~\cite{Chevy:2006}:
\begin{equation}
 |\chi\> = \chi_0|\bm{0}_\parallel\>_A|\mathrm{FS}\>_B 
  + \sum_{|\p|<\hbar\kFB}^{|\q|>\hbar\kFB}\chi_{\p,\q}
  |\p_\parallel-\q_\parallel\>_A|\p,\q\>_B.
\end{equation}
Here $|\p_\parallel\>_A$ is the momentum eigenstate of a single $A$ atom
and $|\mathrm{FS}\>_B$ is the unperturbed Fermi sea of $B$ atoms with
the Fermi momentum $\hbar\kFB=\sqrt{2m_B\mu_B}$.  $|\p,\q\>_B$ describes
the perturbed Fermi sea with the particle (momentum $\q$) and hole
($\p$) excitations.  This simple approximation is known to give
reasonable results in the usual 3D case~\cite{Combescot:2007}.  By
minimizing the energy expectation value with respect to the coefficients
$\chi_0$ and $\chi_{\p,\q}$, we obtain the equation to determine
$\mu_A^\cri$:
\begin{align}
 &\sum_{|\p|<\hbar\kFB}\!
 \biggl\{\sum_\q\,\biggl[\frac{\theta(|\q|-\hbar\kFB)}
 {\frac{(\p_\parallel-\q_\parallel)^2}{2m_A}
 {+}\frac{\q^2-\p^2}{2m_B}{-}\mu_A^\cri}
 - \frac1{\frac{{\q_\parallel}^{\!\!2}}{2m_A}
 {+}\frac{\q^2}{2m_B}}\biggr]\biggr\}^{-1} \notag\\
 &= \mu_A^\cri.   
\end{align}
Similarly, it is straightforward to derive the equation to determine
$\mu_B^\cri$.  Numerically obtained critical chemical potentials
$\mu_A^\cri/\mu_B$ and $\mu_B^\cri/\mu_A$ as functions of the mass ratio
$u_{\min}<u<u_{\max}$ are plotted in Fig.~\ref{fig:chemical_potential}
for the 2D-3D and 1D-3D mixtures.  

\sect{Other universal systems}
Finally we elucidate other possible scale-invariant systems with
short-range few-body interactions.  First it is possible for an
additional intra-species interaction to produce a resonance in a
three-body channel~\cite{Nishida:2007mr}.  $AAB$ three-body resonance
can be introduced if the three-body wave function \eqref{eq:Psi_AAB}
with $\gamma=\gamma_-$ is normalizable at origin $R\to0$.  This
corresponds to the range of mass ratio $2.33<u<6.35$ 
for the 2D-3D mixture or $0<u<2.06$ for the 1D-3D mixture.  Similarly,
$ABB$ three-body resonance can be introduced if the mass ratio is in the
range $0.0351<u<0.0661$ 
for the 2D-3D mixture or $0.00646<u<0.0202$ 
for the 1D-3D mixture.  The corresponding many-body systems become
universal Fermi gases with both two-body and three-body resonances.  

It is also possible to realize scale-invariant systems in different
spatial configurations from those considered so far.  The key idea is to
find a system in which the relative motion is characterized by 
{\em three\/} coordinates (excluding the center-of-mass translations if
exist).  In such a system, the interspecies short-range interaction is
solely characterized by the effective scattering length and at resonance
the scale invariance is achieved.  Consider a 1D-2D (2D-2D) mixture
where $B$ atoms are confined in a 2D plane having a point-like (linear)
intersection with the 1D line (other 2D plane) in which $A$ atoms are
confined.  Here $A$-$B$ relative motions are three-dimensional and the
resulting systems at resonance provide two more classes of universal
two-species Fermi gases in mixed dimensions.

Our idea can be extended to cases with more atomic species.  If the
number of atomic species is three ($A$, $B$, and $C$), there exist two
classes of universal three-species Fermi gases in mixed dimensions.  One
is a 1D$^2$-2D mixture where $A$ and $B$ atoms are confined in the same
1D line embedded in a 2D plane where $C$ atoms are confined.  The other
is a 1D-1D-1D mixture where $A$, $B$, and $C$ atoms are independently
confined in 1D lines and these three lines intersect at a single point.
In either case, the interspecies three-body interaction is
three-dimensional and can be in principle tuned to resonance.

Furthermore, if we have four atomic species and all of them are confined
in the same 1D line, the interspecies four-body interaction becomes
three-dimensional.  The corresponding many-body system at resonance will
be especially interesting because it forms a universal four-species
Fermi gas purely in one dimension.

\sect{Conclusion}
In addition to the well-known $s$-wave two-body resonant interaction in
pure 3D, there exist seven more types of scale-invariant short-range
few-body interactions in pure and mixed dimensions: two-body resonant
interactions in 2D-3D, 1D-3D, 2D-2D, and 1D-2D mixtures; three-body
resonant interactions in 1D$^2$-2D and 1D-1D-1D mixtures; four-body
resonant interaction in pure 1D.  At finite densities, each of them
corresponds to a novel class of universal multispecies Fermi gases that
have nontrivial many-body physics.  We have given detailed analyses in
the cases of two-species Fermi gases in the 2D-3D and 1D-3D mixtures.

\acknowledgments
Discussions with D.~T.~Son are acknowledged.  This work is supported by
JSPS Postdoctoral Program for Research Abroad and by DOE Grant
No.\ DE-FG02-00ER41132.


\begin{thebibliography}{99}
\bibitem{exact_relationship}
If the solution $\gamma$ is real, the wave function at zero energy can
be mapped into an energy eigenstate in a harmonic potential
$V_\omega=\sum_i^{N_A}\frac12m_A\omega^2\x_{Ai}^{\,2}
+\sum_i^{N_B}\frac12m_B\omega^2\x_{Bi}^{\,2}$.  
The corresponding ladder of energy eigenvalues is given by
$E_n=\left(\frac{3-d_\perp}2N_A+\frac32N_B+\gamma+2n\right)\hbar\omega$,
where $n=0,1,2,\ldots$ and $N_{A(B)}$ is a number of $A(B)$ atoms.
\bibitem{footnote_2D}
In the $s$-wave channel, the solutions $\gamma=0,-3$ are forbidden
because they correspond to $\Psi=0$ in real space.
\bibitem{footnote_1D}
In the even-parity channel, the solutions $\gamma=0,-2$ are forbidden
because they correspond to $\Psi=0$ in real space.
\end{thebibliography}
\end{document}